\documentclass[twocolumn,amsmath,amssymb,floatfix,prl,nofootinbib,nolongbibliography]{revtex4-2}
\usepackage{epsfig}
\usepackage{hyperref}
\hypersetup{
           breaklinks=true,   
           colorlinks=false,   
           pdfusetitle=true,  
        }
\begin{document}
\title{NNLO Global Analysis of Polarized Parton Distribution Functions}
%
\author{Ignacio Borsa}  
\email{ignacio.borsa@itp.uni-tuebingen.de}
\affiliation{Institute for Theoretical Physics, University of T\"ubingen, Auf der Morgenstelle 14, 72076 T\"ubingen, Germany}
\author{Daniel de Florian}  
\email{deflo@unsam.edu.ar}
\affiliation{International Center for Advanced Studies (ICAS), ICIFI and ECyT-UNSAM, 25 de Mayo y Francia, (1650) Buenos Aires, Argentina}

\author{Rodolfo Sassot}  
\email{sassot@df.uba.ar}
\affiliation{Departamento de F\'{\i}sica and IFIBA, Facultad de Ciencias Exactas y Naturales, 
Universidad de Buenos Aires, Ciudad Universitaria, Pabell\'on\ 1 (1428) Buenos Aires, 
Argentina}

\author{Marco Stratmann}  
\email{marco.stratmann@uni-tuebingen.de}
\affiliation{Institute for Theoretical Physics, University of T\"ubingen, Auf der Morgenstelle 14, 72076 T\"ubingen, Germany}

\author{Werner Vogelsang}  
\email{werner.vogelsang@uni-tuebingen.de}
\affiliation{Institute for Theoretical Physics, University of T\"ubingen, Auf der Morgenstelle 14, 72076 T\"ubingen, Germany}

\begin{abstract}
We present a next-to-next-to-leading order (NNLO) global QCD analysis of the proton's 
helicity parton distribution functions (PDFs), the first of its kind.
To obtain the distributions, we use data for longitudinal spin asymmetries in inclusive 
and semi-inclusive lepton-nucleon scattering as well as in weak-boson and hadron or jet
production in proton-proton scattering. We analyze the data using QCD perturbation theory at NNLO accuracy, 
employing approximations provided by the threshold resummation formalism
in cases where full NNLO results for partonic hard-scattering functions are not readily available. 
Our numerical results suggest a remarkable perturbative stability of the extracted helicity PDFs. 
\end{abstract}


\maketitle

%
{\it Introduction.---} 
The investigation of the inner workings of the nucleon is paramount in the realm of nuclear and particle physics. 
A focal point lies in unraveling the spin structure of the nucleon, shedding light on the intricate interplay between 
the spins and orbital angular momenta of its constituent quarks and gluons. 
Deep-inelastic scattering (DIS) experiments with polarized leptons and nucleons have revealed that a 
significant fraction of the proton's spin is not
due to the spins of quarks and antiquarks, sparking notable theoretical advancements and new pioneering experiments \cite{RevModPhys.85.655}. 
Among these, polarized proton-proton ($pp$) collisions at the BNL Relativistic Heavy Ion Collider (RHIC) have opened a new frontier,  
providing the first evidence of positive polarization of gluons within the proton \cite{deFlorian:2014yva,Nocera:2014gqa}.

With the advent of the Electron Ion Collider (EIC), expected to commence operations in the early 2030s, 
a new era of precision study of the proton's spin structure will 
unfold, with percent-level accuracy of DIS and semi-inclusive DIS (SIDIS) observables 
expected over an extended kinematical range in Bjorken-$x$ 
and photon virtuality $Q^2$. This prospect requires a match in theoretical precision 
and motivates the move to higher orders in QCD perturbation theory
in the analysis of the helicity PDFs, specifically to NNLO \cite{AbdulKhalek:2021gbh}. 
Not only are the NNLO corrections expected to be numerically relevant in the kinematic
regimes pertinent for extracting helicity PDFs from present and future (EIC) data, 
they are also important for reducing the uncertainty of the
theoretical calculations with respect to changes of factorization and renormalization scales. 
In short, the goal must be to emulate the successful pathway to precision 
taken in studying the unpolarized structure of the nucleon over the last decades. 
Recently, there have been two initial efforts to extract NNLO spin-dependent 
parton distributions. In Ref.~\cite{Taghavi-Shahri:2016idw} the data from polarized inclusive DIS were analyzed at NNLO, 
while Ref.~\cite{Bertone:2024taw} additionally included the available SIDIS data. 
The purpose of our paper is to present the first fully global NNLO analysis, taking into account
the world DIS, SIDIS, and $pp$ spin asymmetry data. As is well known, especially the latter are vital for pinning down 
the gluon helicity PDF, since jet or hadron production at high transverse momentum $p_T$ receive significant contributions 
from gluon-induced hard scattering. 
Therefore, in order to determine a complete and accurate set of polarized PDFs at NNLO, 
it is mandatory to perform a global analysis that includes all the available observables. 

{\it Ingredients to the analysis.---} 
The (anti)quark and gluon helicity PDFs are given by
\begin{equation}
\label{eq:pdf}
\Delta f_j(x,\mu^2) \equiv f^+_j(x,\mu^2) -
                       f^-_j(x,\mu^2),
\end{equation}
where $f^+_j(x,\mu^2)$ [$f^-_j(x,\mu^2)$] 
represents the distribution of a parton of type 
$j$ with positive [negative] helicity in a nucleon with positive helicity, 
at light-cone momentum fraction $x$ and some hard scale $\mu\sim Q$ or $p_T$.
The integral $\int_0^1 \Delta f_j(x,\mu^2) dx$ 
measures the aggregate spin contribution of parton $j$ to the proton spin, which motivates the worldwide efforts 
to extract the $\Delta f_j(x,\mu^2)$ 
through comprehensive fits of data from polarized collisions across various observables. 
Such fits use factorization of cross sections
for hard-scattering reactions and QCD perturbation theory. As an example, for a typical (differential) 
cross section in $pp$ scattering, the factorized
spin-dependent cross section is schematically given by
\begin{equation}\label{eq:2}
d\Delta \sigma^{pp}=\sum_{i,j} \Delta f_i(x_i,\mu)\otimes \Delta f_j(x_j,\mu)\otimes d\Delta \hat{\sigma}_{ij}\,,
\end{equation}
where the $\otimes$ symbol denotes a suitable convolution integral 
and the $\Delta \hat{\sigma}^{ij}$ are spin-dependent partonic 
cross sections for partons $i,j$ producing the observed 
final state\footnote{For observed hadrons in the final state, an additional
convolution with a fragmentation function (FF) will occur.}. 

QCD perturbation theory enters in Eq.~(\ref{eq:2}) in two places. 
The first is the DGLAP scale evolution of the polarized PDFs. The 
corresponding evolution kernels for a parton transition $j\to i$ may be expanded in 
powers of the QCD coupling constant $\alpha_s$ as follows:
\begin{equation}
\Delta {\cal P}_{ij}=\frac{\alpha_s}{2\pi} \, \Delta P^{(0)}_{ij}+\left(\frac{\alpha_s}{2\pi}\right)^2  \Delta P^{(1)}_{ij}+
\left(\frac{\alpha_s}{2\pi}\right)^3  \Delta P^{(2)}_{ij}\,+\ldots,
\end{equation}
where the terms on the right-hand-side successively provide the leading order (LO), 
next-to-leading order (NLO), and NNLO approximations to the kernels. 
The NNLO terms have been computed in Refs.~\cite{Vogt:2008yw,Moch:2014sna,Moch:2015usa,Blumlein:2021enk,Blumlein:2021ryt}, and implemented in a modified version of the {\tt PEGASUS} evolution library \cite{Vogt:2004ns}. The evolution of polarized PDFs up to NNLO was benchmarked against the evolution libraries {\tt EKO} \cite{Candido:2022tld} and {\tt APFEL} \cite{Bertone:2013vaa}, finding excellent agreement.

Likewise, the partonic hard-scattering functions  may be 
evaluated in QCD perturbation theory in terms of an expansion
\begin{equation}
d\Delta \hat{\sigma}_{ij}=d\Delta\hat{\sigma}_{ij}^{\mathrm{(0)}} \,+\,\left(\frac{\alpha_s}{\pi}\right)d\Delta\hat{\sigma}_{ij}^{\mathrm{(1)}} 
\,+\,\left(\frac{\alpha_s}{\pi}\right)^2d\Delta\hat{\sigma}_{ij}^{\mathrm{(2)}} \,+\,\ldots\,.
\end{equation}
For a NNLO analysis, knowledge of the $d\Delta\hat{\sigma}_{ij}^{\mathrm{(2)}}$ is required.
The corresponding corrections to polarized inclusive DIS have been known for a long time \cite{Zijlstra:1993sh}, 
while the results for SIDIS became available only very recently 
\cite{Goyal:2023zdi,Goyal:2024tmo,Bonino:2024qbh,Bonino:2024wgg}\footnote{
We note that the NNLO corrections are also known for the full set of inclusive DIS electroweak polarized structure functions \cite{Borsa:2022irn} and for jet production in DIS \cite{Borsa:2022cap}, albeit both unmeasured so far.}. 
Among the processes relevant at RHIC so far only the NNLO corrections to $pp\to \ell X$
via electroweak-boson production and decay are available \cite{Boughezal:2021wjw} 
(see also earlier work on the Drell-Yan process in \cite{Ravindran:2003gi}),
but not yet those for the all-important probes of gluon polarization, $pp\to \textrm{jet}\,X$ and $pp\to hX$. 

In order to fill this gap and be able to do a fully global NNLO analysis, 
we resort to the use of approximate NNLO expressions
for the observables for which the full set of NNLO corrections is not readily available. 
This is the case for the high-$p_T$ 
reactions in $pp$ scattering, but also for SIDIS, where the full numerical evaluation of the corrections presented 
in \cite{Goyal:2023zdi,Goyal:2024tmo,Bonino:2024qbh,Bonino:2024wgg} 
is currently still ongoing and not yet at a level that makes inclusion in a global analysis feasible.
The approximations we use for the NNLO corrections are derived from soft-gluon threshold resummation.
As an example, in SIDIS, the partonic cross sections exhibit large double-logarithmic corrections at large values of 
the partonic variables $x=Q^2/(2p\cdot q)$ and $z=p\cdot p'/p\cdot q$, where $p$ and $p'$ are the momenta
of the incoming and fragmenting quark in the SIDIS reaction, respectively, and $q$ is the virtual photon's momentum. 
Using Mellin moments $N$ and $M$ conjugate to $x$ and $z$, the leading double-logarithmic corrections take the form
$\alpha_s^k \big(\ln N+\ln M)^{2k}\equiv \alpha_s^kL^{2k}$ at the $k$th order of perturbation theory. 
As is well known, these terms may be resummed to all orders in perturbation theory 
\cite{Cacciari:2001cw,Sterman:2006hu,Anderle:2012rq,Abele:2021nyo},
giving rise to an exponentiated form of the all-order cross section near partonic threshold:
\begin{equation}
\Delta \hat{\sigma}_{qq}\propto\exp\left[ Lh_1(\alpha_sL)+h_2(\alpha_sL)+\alpha_sh_3(\alpha_sL)+\ldots\right],
\end{equation}
where the functions $h_i$ are single-logarithmic. Expanding this expression to ${\cal O}(\alpha_s^2)$ gives
the desired approximate soft-gluon NNLO corrections for SIDIS. They turn out to be identical for the spin-averaged and
spin-dependent cases. All details relevant at next-to-next-to-leading logarithmic accuracy 
may be found in \cite{Abele:2021nyo}. For the processes $pp\to \textrm{jet}\,X$ and $pp\to hX$
similar techniques may be used, although here the formalism is more involved since the underlying LO partonic 
processes involve four color-charged particles, giving rise to a matrix problem for the resummed cross section \cite{Kidonakis:1997gm,Kidonakis:2000gi,Catani:2013vaa}
in which soft and hard contributions are coupled via a trace in color space. As a result, one obtains different resummed
expressions in the polarized and unpolarized cases. Also, the kinematics of single-particle or jet production 
affects the form of the resummed cross section. 
Details may be found in Refs.~\cite{Hinderer:2018nkb,deFlorian:2007tye,Hinderer:2017bya}.
We note that we apply the soft-gluon approximations only to the NNLO terms in the cross sections but
keep the NLO part in full without approximation. Technically, the NNLO terms are included in terms of $K$-factors.
For the inclusive-DIS and $W^{\pm}$-boson production processes, we incorporate the full NNLO results. 
We note that we do not use the RHIC dijet data in our analysis since the approximate NNLO framework for dijets
is not yet available. 

{\it Fitting methodology.---} 
The technique for our global analysis has been detailed in Refs.~\cite{deFlorian:2009vb,DeFlorian:2019xxt} 
and will not be reiterated here. It relies on an efficient Mellin-moment technique, 
enabling the tabulation and storage of the most computationally demanding components 
of the calculation before the actual analysis. Consequently, the computation of the pertinent 
spin-dependent $pp$ cross sections is expedited to such an extent 
that it can be seamlessly incorporated within a standard $\chi^2$-minimization analysis.
It is this step, that is still under development for replacing the approximate by the
exact but numerically demanding NNLO expressions for SIDIS.

For our NLO and NNLO global fits we adopt the same flexible functional form as in~\cite{deFlorian:2009vb} for 
the helicity parton densities of light-quarks and gluon at the initial scale $Q_0=1$~GeV, for instance,
%
%
\begin{equation}
x\Delta f_i(x,Q_0^2)=N_i x^{\alpha_i}(1-x)^{\beta_i}\left(1+\gamma_i x+\eta_i x^{\kappa_i}\right),
\label{dginp}
\end{equation}
with free parameters $N_i$, $\alpha_i$, $\beta_i$, $\gamma_i$, $\eta_i$, and $\kappa_i$. 
We enforce positivity $|\Delta f|/f\leq 1$ of the parton densities, 
using the unpolarized distributions $f(x,Q^2)$ of \cite{Bailey:2020ooq}, 
from which we also adopt the running of the strong coupling at each corresponding order, and which we also use
for the spin-averaged cross sections in the denominators of the measured spin asymmetries.
The distributions corresponding to heavy-quarks are generated perturbatively above their mass thresholds in the zero-mass variable-flavor-number scheme, including the corresponding matching coefficients 
at threshold from \cite{Bierenbaum:2022biv}. We do not consider heavy-quark masses effects, which are expected to be small compared to the current experimental uncertainties, although they will play a more prominent role in the future EIC \cite{Hekhorn:2024tqm}. 
 
For processes involving observed hadrons in the final state, we need a set of fragmentation functions (FFs). 
As no \textit{full} global analysis of FFs is available at 
NNLO accuracy\footnote{NNLO analyses of FFs using only $e^+e^-$ 
and SIDIS data do exist \cite{Borsa:2022vvp,AbdulKhalek:2022laj}. 
Unfortunately, these data sets do not allow for a proper extraction of the gluon FF, which is essential for the theoretical description of $pp\to hX$ results.}, 
we have to rely on the most comprehensive NLO fits from \cite{Borsa:2021ran,Borsa:2023zxk}. 
While some effects of the FFs cancel in the spin asymmetries, 
this choice introduces an extra source of theoretical uncertainty in our NNLO result.  
Additional challenges arise in the description of the available SIDIS data at $x\lesssim 0.1$, 
where the associated values of $Q^2$ are below the regime of validity of the unpolarized
baseline PDFs and where SIDIS is not well described by current FFs \cite{Borsa:2021ran}. 
Furthermore, in this regime the full NNLO corrections to SIDIS appear
to be especially sizable \cite{Bonino:2024wgg} and are not represented well by the approximate NNLO results.
Indeed, we have found that fits that include the SIDIS spin asymmetry data at $x\lesssim 0.1$ do not 
work particularly well and show a significant decrease in quality 
when going from NLO to NNLO, consistent with observations made in Ref.~\cite{Bertone:2024taw}.
Therefore, in order to be conservative and to ensure a robust analysis, 
we apply a cut $x > 0.12$ to the SIDIS data included in the fit, which results in
comparable $\chi^2$-values for SIDIS at NLO and NNLO accuracy in our analysis. 
Commensurate with this cut, we also apply a cut 
of $p_T>1.5$ GeV for the RHIC $pp \rightarrow hX$ data. 

In order to assess the estimated residual uncertainties of the extracted polarized PDFs, 
we determine a set of 600 PDF Monte Carlo sampling replicas \cite{DeFlorian:2019xxt}.
These sets may be straightforwardly used to propagate the PDF uncertainty to other spin observables. 

\begin{figure}[th!]
\epsfig{figure=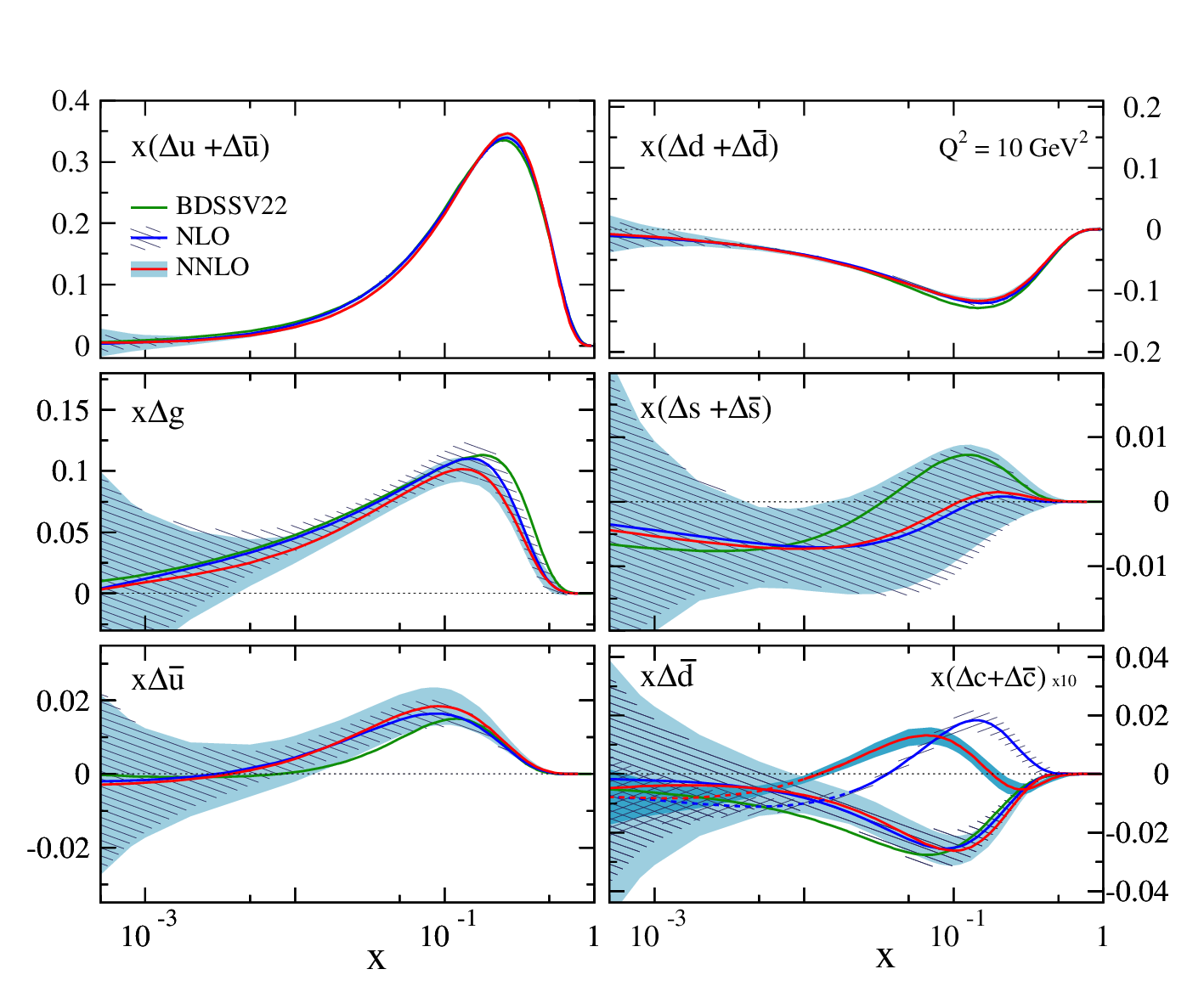,width=0.5\textwidth}

\vspace*{-0.4cm}
\caption{Our helicity PDFs obtained from global analysis at NLO and NNLO accuracy, 
along with their uncertainty bands at $68\%$ confidence level.
Also shown is the previous BDSSV22
fit at NLO, see text.}\label{fig1}
\end{figure}

{\it Results for NLO and NNLO polarized PDFs.---} 
Figure~\ref{fig1} showcases our newly obtained NLO (blue lines) and NNLO (red lines) 
BDSSV24 distributions alongside their respective uncertainty bands at $Q^2=10\,\mathrm{GeV}^2$. 
%

A remarkable perturbative stability is observed, with minimal changes of the PDFs from NLO to NNLO both in terms of their central values and uncertainties.
This is also reflected in the total $\chi^2$--values of the fit after application of the cuts, which are reported in Table \ref{table}.  
The inclusion of NNLO corrections leads to improvements, albeit small, across all sets of data 
with the exception of $pp\to hX$, where $\chi^2$ remains essentially unchanged.

\begin{table}[b]
\caption{\label{table} Partial and total $\chi^2$ obtained in the fits}
\begin{tabular}{lcccccc}
\hline \hline
  & Total & DIS & SIDIS &  pp-jets & pp-pions  & pp-W \\
\hline
NLO & 627.2 & 302.7  &  127.6  & 111.1 & 63.5     & 22.3\\
NNLO & 607.5 & 294.3  & 122.9  & 104.0 & 66.0  & 20.3 \\ \hline
Data points & 673    &  368  & 114  &  91   & 78    & 22  \\ \hline
\end{tabular}\end{table}

As expected, the total up and down quark PDFs are well determined. 
The gluon helicity PDF is seen to be clearly positive, with relatively small uncertainty, over a 
broad region in $x$ accessed by RHIC data. Furthermore, there is a pronounced pattern of positive 
$\Delta\bar u$ and negative $\Delta\bar d$ at moderately large $x$, while the strangeness helicity PDF 
is essentially undetermined. This is in contrast to our previous analyses \cite{deFlorian:2009vb,deFlorian:2014yva}, 
and results from the fact that we no longer impose constraints related to 
SU(2) and SU(3) symmetry\footnote{Nevertheless, the 
central NLO and NNLO fits both deviate only by a few per cent 
from these constraints that are usually given in terms of $F$ and $D$ values.}, as well as from the reduced constraints from SIDIS data due to the use of an $x$-cut. 
Sizable uncertainties remain for the sea quark and gluon PDFs
at $x\lesssim 10^{-3}$, consistent with the lack of experimental data constraining that region.

\begin{figure}[th!]
\vspace*{-0.75cm}

\epsfig{figure=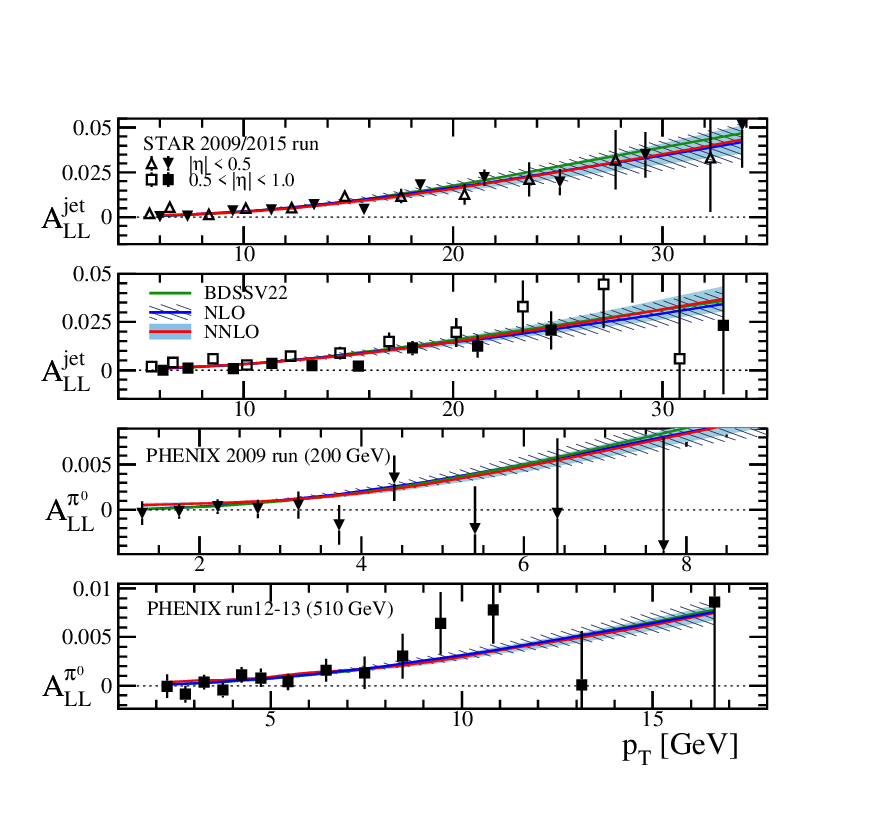,width=0.53\textwidth}

\vspace*{-0.5cm}
\caption{Comparison of our fit results at NLO and NNLO accuracy to RHIC double-spin asymmetries 
for $pp\to hX$ and $pp\to\textrm{jet}X$ \cite{RHICSPIN:2023zxx}.}\label{fig2}

\vspace*{-0.5cm}
\epsfig{figure=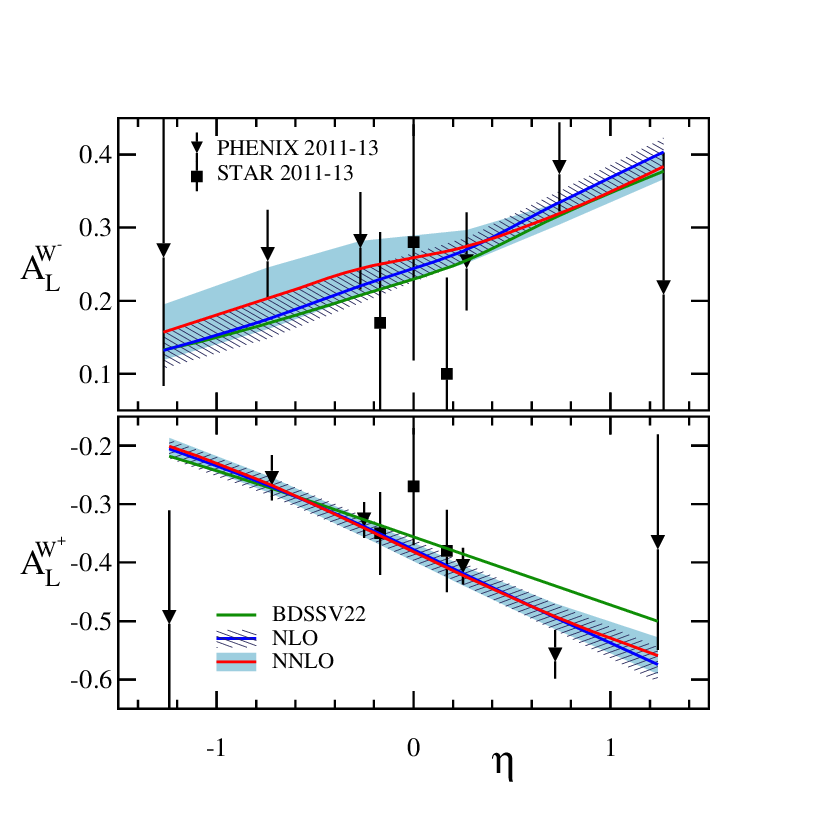,width=0.53\textwidth}

\vspace*{-0.5cm}
\caption{Same as in Fig.~\ref{fig2} but now for the
RHIC single-spin asymmetries for $W^\pm$-boson production \cite{RHICSPIN:2023zxx}.}\label{fig3}
\vspace*{-0.5cm}
\end{figure}

\begin{figure}[th!]
\vspace*{-0.5cm}
\epsfig{figure=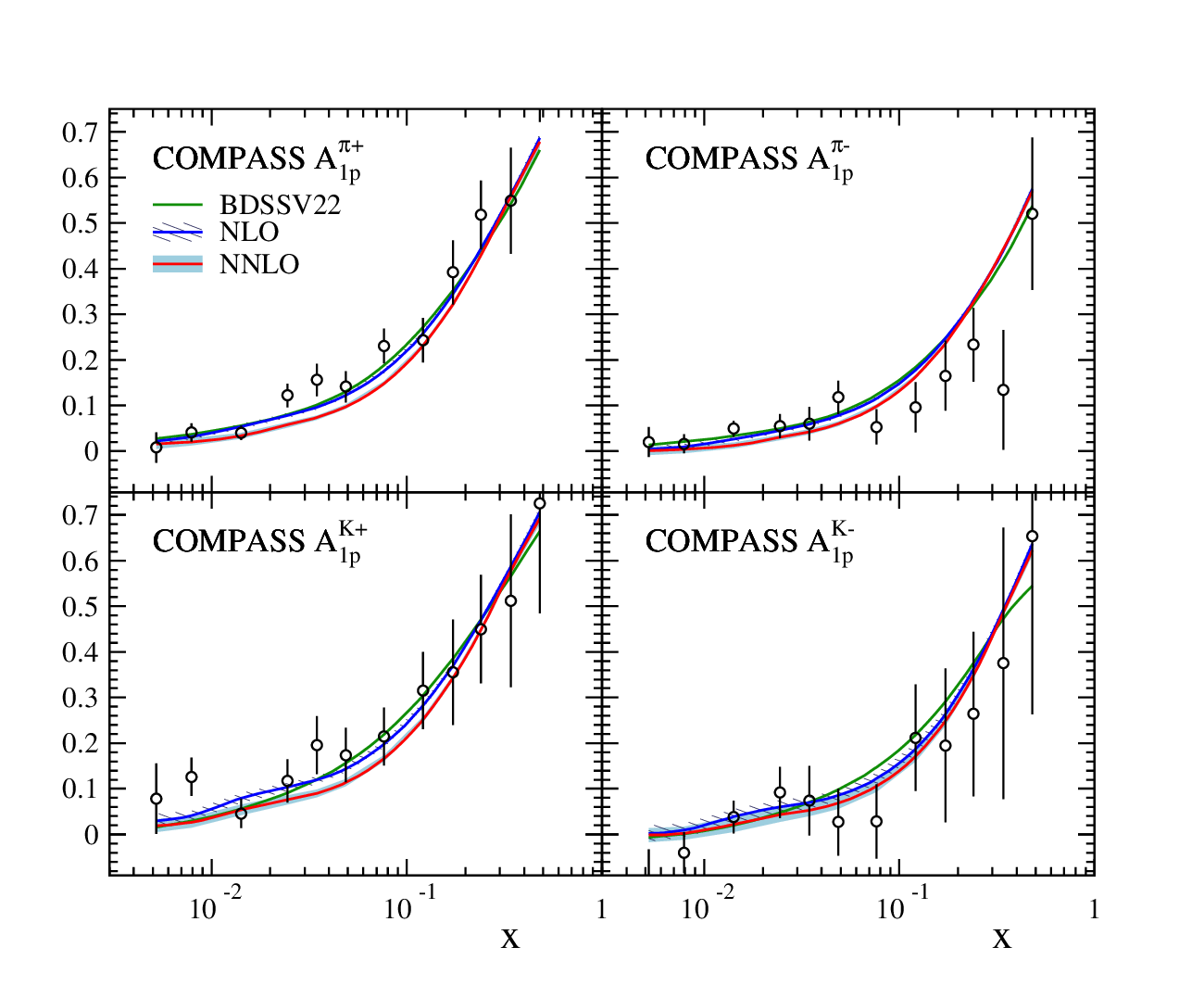,width=0.50\textwidth}

\vspace*{-0.5cm}
\caption{Same as in Fig.~\ref{fig2} but now for selected SIDIS spin asymmetry data for pion and kaon production
from the COMPASS experiment \cite{COMPASS:2016xvm}.}\label{fig4}

\vspace*{-0.5cm}
\epsfig{figure=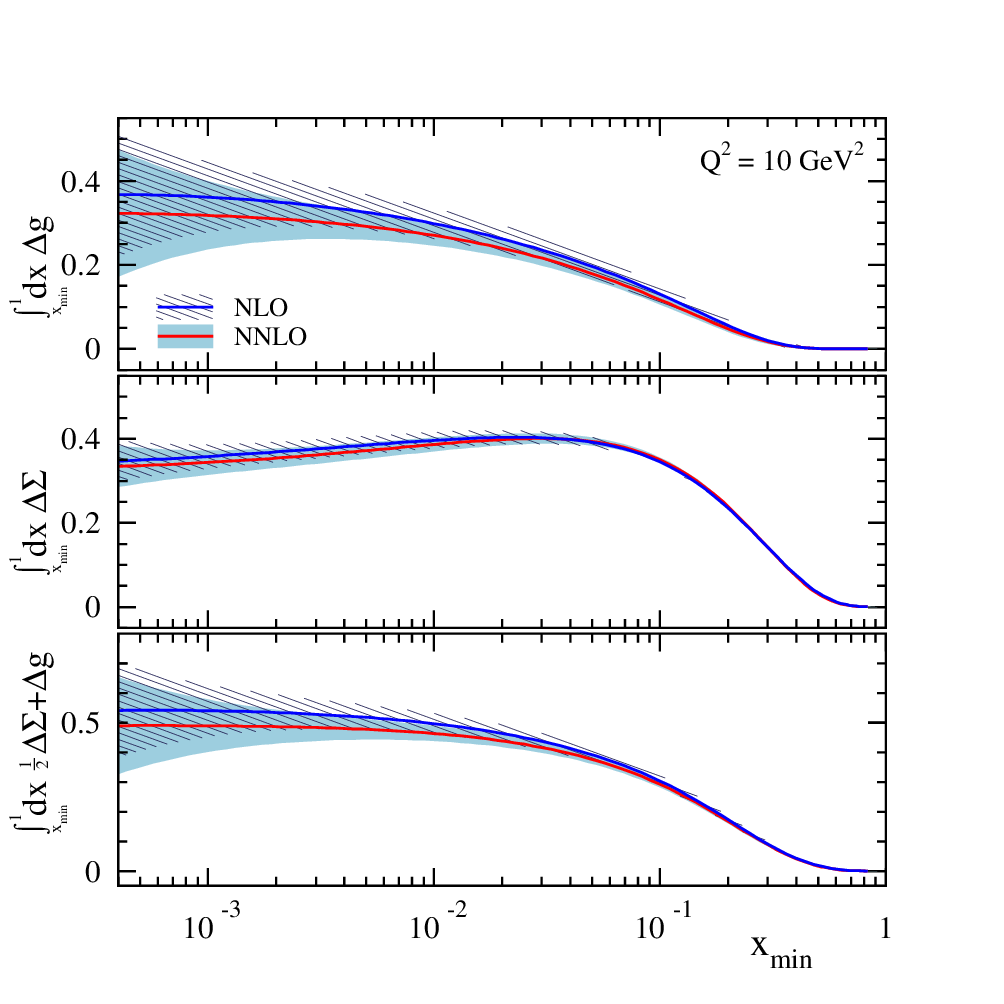,width=0.50\textwidth}

\vspace*{-0.5cm}
\caption{Truncated first moments of the quark singlet and gluon helicity PDFs at $Q^2=10$~GeV$^2$, and their combined
contribution to the proton spin sum rule (bottom panel).}\label{fig5}
\vspace*{-0.5cm}
\end{figure}

Figures~\ref{fig2}--\ref{fig4} show detailed comparisons of our fit results at NLO and NNLO accuracy
to a few selected sets of $pp$ and SIDIS data that were used in our analysis. 
For comparison, we also show in Figs.~\ref{fig1}--\ref{fig4}
the results for an (unpublished) NLO fit (labeled as BDSSV22) 
obtained using the same methodology of \cite{deFlorian:2014yva}, but updating the data sets to additionally include RHIC dijet and $W^{\pm}$ production data (for a collection of RHIC data, see, e.g.\ \cite{RHICSPIN:2023zxx}), as well as COMPASS final SIDIS measurements \cite{COMPASS:2016xvm}. 
As can be seen, the differences to our new NLO fit are rather moderate and mostly within the
estimated uncertainty bands.

We finally use our extracted NLO and NNLO distributions to explore the quark and gluon spin contributions 
to the proton spin. While we cannot reliably determine the full first moments due to the sizable 
extrapolation uncertainties at low $x$, we can present results for the truncated moments. 
This is done in Fig.~\ref{fig5} for the quark singlet $\int_{x_\textrm{min}}^1 dx\,\Delta \Sigma$ and
for the gluon, $\int_{x_\textrm{min}}^1 dx\,\Delta g$, as functions of the lower integration limit $x_\textrm{min}$, 
at $Q^2=10$~GeV$^2$. 
A striking feature of the results shown is the relatively flat behavior towards low $x_\textrm{min}$, 
suggesting good convergence of the integrals. 
Remarkably, when combining the two contributions according to their role for the proton spin, 
one finds a result approaching 1/2 toward lower $x_\textrm{min}$. It will be interesting to see 
whether future data indeed confirm this indication of a small
contribution by orbital angular momenta to the proton spin.

{\it Conclusions.---} We have carried out the first global analysis of helicity PDFs at the NNLO of QCD. 
The NNLO corrections to spin-dependent partonic cross sections and to PDF evolution available in the literature
have been augmented by results derived from soft-gluon threshold resummation where necessary, resulting
in a framework that fully incorporates at least approximate NNLO corrections. 
We observe a remarkable perturbative stability of the extracted helicity PDFs from NLO to NNLO. 

Our study sets the stage for future NNLO global analyses of nucleon helicity structure in the EIC era. 
There are several avenues for future improvements, among them derivations of better approximations for the NNLO terms
in $pp$ scattering, the implementation of the full NNLO results for SIDIS, detailed explorations of the residual scale dependence and power corrections and the inclusion of theoretical uncertainties in the fit \cite{NNPDF:2019vjt,NNPDF:2019ubu,NNPDF:2024dpb}.

Both the NLO and NNLO sets of replicas are publicly available in LHAPDF format in \url{https://github.com/igborsa/BDSSV24}.

\acknowledgments
We thank R. Thorne for providing the NNLO code for the DIS structure functions of
Ref.~\cite{Bailey:2020ooq}, R. Boughezal for the NNLO corrections for $W^{\pm}$ production \cite{Boughezal:2021wjw}, and L. Bonino for the NNLO SIDIS coefficient functions from 
\cite{Bonino:2024qbh,Bonino:2024wgg}.
This work was partially supported by CONICET and ANPCyT and by Deutsche Forschungsgemeinschaft 
(Research Unit FOR 2926, project 409651613).
\bibliography{bdssv24-post-review}
\end{document}